\documentclass[letterpaper,prl,superscriptaddress,floatfix,twocolumn,showpacs,preprintnumbers,amsmath,amssymb]{revtex4}

\usepackage{graphicx}
\usepackage{dcolumn}
\usepackage{bm}
\newcommand{\scalefig}{.85}
\newcommand{\Ne}[1]{\ensuremath{^{ #1}\mathrm{Ne}}}
\newcommand{\Pb}[1]{\ifnum#1>0$^{#1}$Pb\else$^\mathrm{nat}$Pb\fi}
\newcommand{\efm}[1]{\newcount\efmpow
\efmpow=#1
\multiply\efmpow by 2
\ensuremath{\mathrm{e}^2\mathrm{fm}^{\number\efmpow}}}

\newcommand{\Ar}{$^{ 40}$Ar}
\newcommand{\Be}{$^{ 9}$Be}
\newcommand{\Al}[1]{\ifnum#1>0$^{#1}$Al\else$^\mathrm{nat}$Al\fi}
\newcommand{\um}{\ensuremath{\mu\textrm{m}}}
\newcommand{\via}{\textit{via }}

\begin{document}

\preprint{}

\title{Decay Pattern of Pygmy States Observed in Neutron-Rich \Ne{26}}

\newcommand{\ipno}{\affiliation{Institut de Physique Nucl\'eaire,
    IN$_2$P$_3$-CNRS, Universit\'e Paris Sud, F-91406 Orsay, France}}
\newcommand{\rikkyo}{\affiliation{Department of Physics, Rikkyo
    University, 3-34-1 Nishi-Ikebukuro, Toshima, Tokyo 171-8501,
    Japan}} \newcommand{\riken}{\affiliation{RIKEN (The Institute of
    Physical and Chemical Research), 2-1 Hirosawa, Wako, Saitama
    351-0198, Japan}} \newcommand{\atomki}{\affiliation{Institute of
    Nuclear Research of the HungarianAcademy of Sciences, PO Box 51,
    H-4001 Debrecen, Hungary}}
\newcommand{\titech}{\affiliation{Department of Physics, Tokyo
    Institute of Technology, Tokyo 152-8551, Japan}}
\newcommand{\cns}{\affiliation{Center for Nuclear Study, University of
    Tokyo, RIKEN Campus, 2-1 Hirosawa, Wako, Saitama 351-0198, Japan}}
\newcommand{\kyushu}{\affiliation{Department of Physics, Kyushu
    University, 6-10-1 Hakozaki, Higashi, Fukuoka 812-8581, Japan.}}

\author{J.~Gibelin}\email{gibelin@lpccaen.in2p3.fr}\altaffiliation{present address: LPC Caen, Universit\'e de Caen, F-14050 Caen Cedex, France}\ipno\rikkyo

\author{D.~Beaumel}\ipno
\author{T.~Motobayashi}\riken
\author{Y.~Blumenfeld}\ipno
\author{N.~Aoi}\riken
\author{H.~Baba}\riken
\author{Z.~Elekes}\atomki
\author{S.~Fortier}\ipno
\author{N.~Frascaria}\ipno
\author{N.~Fukuda}\riken
\author{T.~Gomi}\riken
\author{K.~Ishikawa}\titech
\author{Y.~Kondo}\titech
\author{T.~Kubo}\riken
\author{V.~Lima}\ipno
\author{T.~Nakamura}\titech
\author{A.~Saito}\cns
\author{Y.~Satou}\titech
\author{J.-A.~Scarpaci}\ipno
\author{E.~Takeshita}\rikkyo
\author{S.~Takeuchi}\riken
\author{T.~Teranishi}\kyushu
\author{Y.~Togano}\rikkyo
\author{A.~M.~Vinodkumar}\titech
\author{Y.~Yanagisawa}\riken
\author{K.~Yoshida}\riken

\date{\today}
             
\begin{abstract}
  Coulomb excitation of the exotic neutron-rich nucleus \Ne{26} on a
  \Pb{208} target was measured at 58~MeV/u in order to search for
  low-lying E1 strength above the neutron emission threshold.
  This radioactive beam experiment was carried out at the RIKEN
  Accelerator Research Facility.
  Using the invariant mass method in the \Ne{25}$+n$ channel, we
  observe a sizable amount of E1 strength between 6 and 10~MeV
  excitation energy.  By performing a multipole decomposition of the
  differential cross-section, a reduced dipole transition probability
  of B(E1)=$0.49\pm0.16~\efm{1}$ is deduced, corresponding to
  4.9$\pm$1.6\% of the Thomas-Reiche-Kuhn sum rule.  For the first
  time, the decay pattern of low-lying strength in a neutron-rich
  nucleus is measured. The extracted decay pattern is not consistent
  with several mean field theory descriptions of the pygmy states.
\end{abstract}



\maketitle

The advent of beams of atomic nuclei with large neutron/proton ratios
has offered the possibility to investigate new phenomena associated
with the excess of neutrons. An often quoted property of such exotic
nuclei is the halo effect, an abnormal extension of matter
distribution observed for the first time in light neutron rich
isotopes in the mid 80's \cite{tanihata:halo}.  Beyond static
properties, the question of the occurrence of new dynamical modes
associated with the excess neutrons has been investigated both
theoretically and experimentally. Predictions in favour of such modes
have been given in the early 90's \cite{suzuki:pdg,isacker:pdg}.  In
these calculations, the dipole response of neutron-rich (n-rich)
nuclei exhibits a small component at energies lower than the standard
Giant Dipole Resonance (GDR), often depicted as the oscillation of a
deeply bound core against a neutron halo or skin, giving rise to a
so-called pygmy resonance. Such modifications of the response function
of nuclei have direct implications for astrophysics. The strong
influence of an -- even small -- percentage of E1 strength located
above particle threshold on neutron capture reactions has been studied
in \cite{Goriely:2004}. More recently, the link between pygmy dipole
strength, neutron skin thickness and symmetry energy in asymmetric
nuclear matter, which has a strong impact on several neutron-star
properties has been stressed \cite{Klim:2007:nskin}.  Experimentally,
the presence of low-lying dipole strength exhausting a sizable amount
of the Thomas-Reiche-Kuhn (TRK) energy weighted sum rule (EWSR) in
n-rich nuclei is now established. It was first revealed in light
drip-line nuclei in breakup reactions using high-Z targets
\cite{tanihata:ppnp}. Later on, the non-resonant nature of the dipole
strength found in some light n-rich nuclei such as $^{11}$Be was
stated \cite{Fukuda:2004ty}. In heavier nuclei, low-lying dipole
strength has been recently observed in Coulomb breakup experiments at
high energy performed at GSI on Oxygen \cite{leisten:oxygens:gdr} and
Tin \cite{adrich:sn:pdr} isotopes. In the latter case, an amount of
nearly 5\% of the TRK sum rule has been measured at around 10~MeV
excitation energy in $^{130,132}$Sn nuclei, in agreement with several
mean field models. Interestingly, conflicting interpretations are
provided by the quoted models concerning the microscopic structure of
these states. Within the relativistic quasi-particle random phase
approximation (QRPA) calculations \cite{paar:prc67}, relatively
collective pygmy states are predicted, while non-relativistic QRPA
including phonon coupling involves essentially individual transitions
\cite{Sarchi04:dipole}.  No conclusion can be brought on the
microscopic structure of these states in the absence of other
observables than the strength distribution. Both approaches
nevertheless agree that the excitations are driven by the excess
neutrons.

In the present work we investigate low-lying dipole strength in the
\Ne{26} isotope for which an important redistribution of dipole
strength as compared to the stable \Ne{20} is predicted by Cao and Ma
\cite{ma:prc}. In this calculation, almost 5\% of the TRK sum rule is
exhausted by a structure centered around 8.5 MeV. This region in
energy is located between the one-neutron and the two-neutron emission
threshold. We performed a Coulomb excitation experiment by bombarding
a lead target by \Ne{26} at intermediate energy, and used the
invariant mass method to reconstruct the B(E1) strength from the
\Ne{26}$\to$\Ne{25}$+n$ channel. After determining the strength
distribution, we extract for the first time neutron branching ratios
of the populated pygmy states to levels in the daughter nucleus
(\Ne{25}). These observables provide detailed insight on the
microscopic structure of the populated states, as long as the observed
decay mode is not statistical \cite{vdw:book}.


The experiment was performed at the RIKEN Accelerator Research
Facility. A secondary \Ne{26} beam was produced through fragmentation
of a 95~MeV/u, 60~pnA \Ar\ primary beam on a 2-mm-thick \Be\
target. The \Ne{26} fragments were separated by the RIKEN Projectile
Fragment Separator (RIPS) \cite{kubo:rips}. Beam particle
identification was unambiguously performed by means of the
time-of-flight (TOF) between the production target and the second
focal plane. The 80\% pure \Ne{26} beam of intensity $\sim5\times10^3$
pps and incident energy 58~MeV/u, was tracked with two parallel-plate
avalanche counters providing incident angle and hit position on the
reaction targets (alternatively 230~mg/cm$^2$ \Pb{0} and 130 mg/cm$^2$
\Al{27}).  Data obtained with the \Al{27}\ target are used in the
following to estimate the contribution of nuclear excitation to the
data.

The outgoing charged fragments were detected using a set of telescopes
placed at 1.2~m downstream of the target. They consisted of two layers
(X and Y) of 500~\um\ single-sided silicon strip detectors (SSD) with
5~mm strips which yielded an energy-loss resolution for neon isotopes
of 1.5~MeV (FWHM). An additional layer used 3~mm-thick Si(Li)
detectors made at the Institut de Physique Nucl\'eaire d'Orsay. The
resolution on the remaining energy (E) was 9~MeV (FWHM). Unambiguous
mass and charge identification of all projectile-like fragments was
obtained using the E--$\Delta$E method.

In-beam gamma rays were detected using the 4$\pi$ gamma array DALI2
\cite{takesato:dali2}, which consists of 152 NaI(Tl) detectors placed
around the target. For 1.3~MeV gamma-rays, the measured efficiency is
approximately 15\% and the energy resolution is 7\% (FWHM).  The
Doppler corrected gamma energy distribution obtained in coincidence
with the \Ne{25} isotope allows us to identify the gamma decay from
the adopted 1702.7(7), 2030(50), and 3316.4(11)~keV excited states.

The hodoscope for neutron detection was an array of 4 layers of 29
plastic rods each, placed 3.5~m downstream of the target.
The total intrinsic efficiency for the detection of 60~MeV
neutrons was calculated to be ~25\%~\cite{fukuda:thesis}. Finally, 29
thin plastic scintillators covered the front face of the wall in order
to veto charged particles as well as to provide an active beam
stopper. The neutron position was determined with an error of
$\pm$3~cm and the energy, from TOF information, with a 2.5~MeV (FWHM)
resolution for the neutrons of interest.


A simulation of the experimental setup using the Geant~3
package~\cite{brun:geant3} was performed in order to correct the data
for the experimental acceptance.  Using this simulation, the angular
distribution for elastic scattering of \Ne{26} on \Pb{0} at 55~MeV/u
was obtained and compared with optical model calculations based on an
optical potential for the \Ne{20}+\Pb{0} system at
40~MeV/u~\cite{beaumel:ne20}.  Another check for both the simulation
and the optical potential used was provided by the \Ne{26}
B(E2;$0_1^+\!\to\!2_1^+$)~\cite{gibelin:26ne:be2:}, extracted by
comparing the shape and the amplitude of the angular distribution of
the experimental inelastic scattering with the corresponding
theoretical calculation.  For the \Ne{26}+\Al{27} reaction, we
empirically generated optical potential parameters \cite{gibelin:phd}
and compared the result with the experimental elastic scattering.

Using the invariant mass method, the excitation energy of an unbound
state in the $^\mathrm{A}\mathrm{X}$ nucleus decaying to a state in
$^\mathrm{A-1}\mathrm{X}$ can be expressed by: $E^* = E_\mathrm{rel} +
S_n + {\textstyle\sum_i}\,E_{\gamma_{i}}$, where $ E_\mathrm{rel}$ is
the relative energy between the neutron and the fragment
$^\mathrm{A-1}\mathrm{X}$, $S_n$ the one neutron emission threshold,
and $\sum_i{E_i}$ the summed energy of the gammas involved in the
subsequent decay of the daughter nucleus
$^\mathrm{A-1}\mathrm{X}$. The gamma detection efficiency was not high
enough to perform an event-by-event gamma calorimetry. Hence, our
reconstruction technique took into account independently the
population of every state of the $^\mathrm{A-1}\mathrm{X}$ daughter
nucleus \cite{gibelin:phd,gibelin:comex02}. This method has been
successfully tested on simulations. We also included the detector
resolutions and hence estimated the excitation energy resolution to be
0.8~MeV at $E^* = 8$~MeV.

\begin{figure}[!ht]
\centering \includegraphics[width=\scalefig\columnwidth]{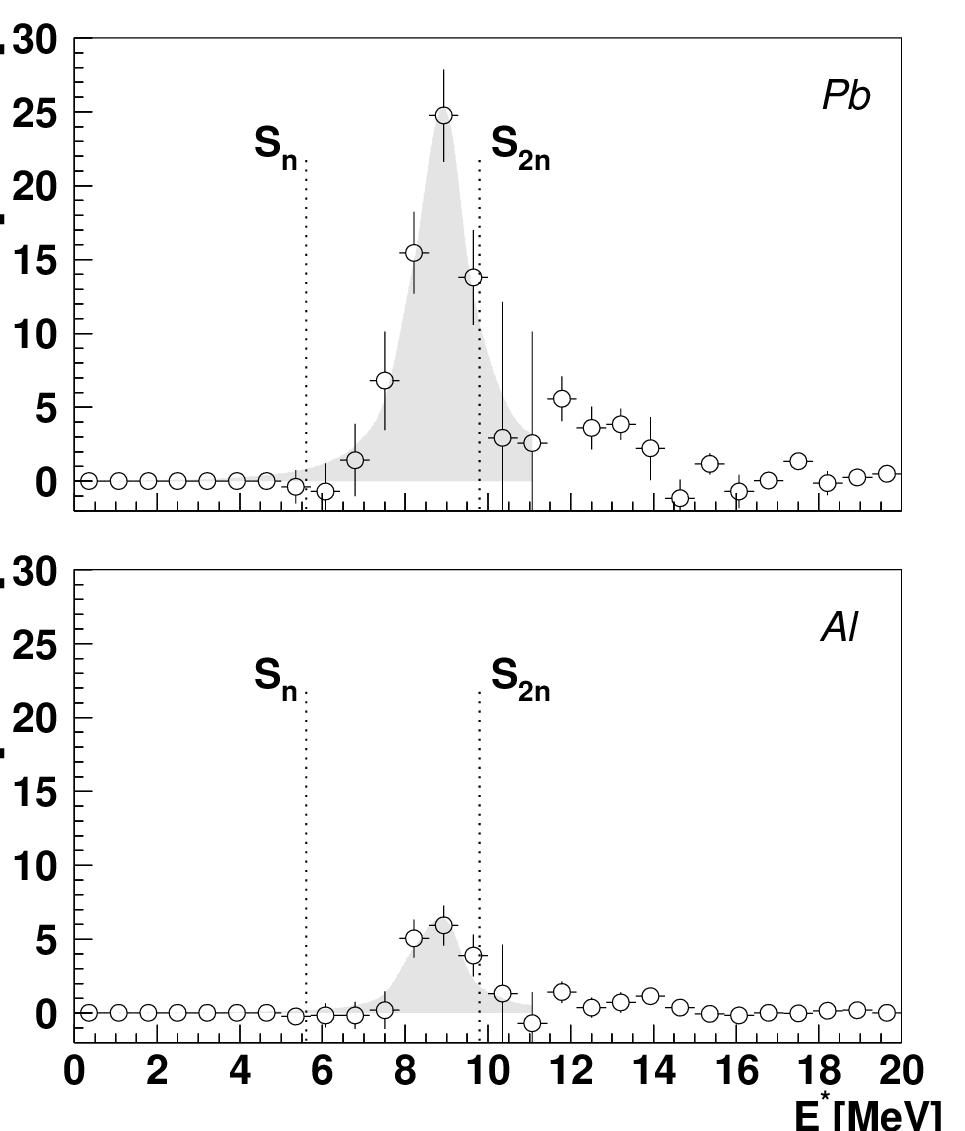}
\caption{\label{fig:invmas}{\bf Top:} Excitation energy distribution
in \Ne{26} reconstructed from the \Ne{25}+n decay channel with
\Pb{0}\ target.  The shaded area is a tentative Lorentzian fit. {\bf
Bottom:} Same as previous but for the \Al{27}\ target.}
\end{figure}

The excitation energy spectra reconstructed for the \Ne{25}$+n$ decay
channel obtained with the Pb and Al targets are represented in
Fig.~\ref{fig:invmas}. Above 10~MeV, the decay of \Ne{26} is expected
to occur mainly by 2-neutron emission. Between 8 and 10~MeV, a sizable
amount of cross-section is observed for both targets. In intermediate
energy inelastic scattering with a heavy target such as Pb, the
spectrum is dominated by Coulomb excitation of E1 states. Conversely,
the dipole excitation is relatively low with a light target such as
Al.  The contribution of possible E2 excitation to the spectrum
obtained with the lead target has been determined by using data taken
with the aluminum target and the coupled channels {\sc ecis97}
code~\cite{ecis:97}.  Assuming a simple collective vibrational mode
with equal nuclear and Coulomb deformation lengths, the E2 nuclear and
Coulomb deformation parameters were extracted from the measured
cross-section with the Al target ($\sigma_\mathrm{Al} =
9.1\pm2.3$~mb). The corresponding $L=2$ cross section in lead was then
calculated using the deformation lengths extracted in the previous
step.
After subtraction of this E2 contribution, the resulting
$\sigma_\mathrm{Pb}^{L=1}=48.5\pm4.8$~mb cross section corresponds to
a Coulomb deformation parameter $\beta_\mathrm{C}=0.087\pm0.008$ which
leads to B(E1)$=0.55\pm0.05$~\efm{1} \via the following relation with
the Coulomb radius $R_C$:
B(E1;$0^+\!\to\!1^-$)=$\left(\frac{3}{4\pi}Z_p\,e\,R_C\beta^{L=1}_C\right)^2$,
where $Z_p$ is the projectile proton number.  This value of reduced
transition probability corresponds to 5.5$\pm$0.6\% of the TRK sum
rule for an excitation energy of 9~MeV.

\begin{figure}[!ht]
\includegraphics[width=\scalefig\columnwidth]{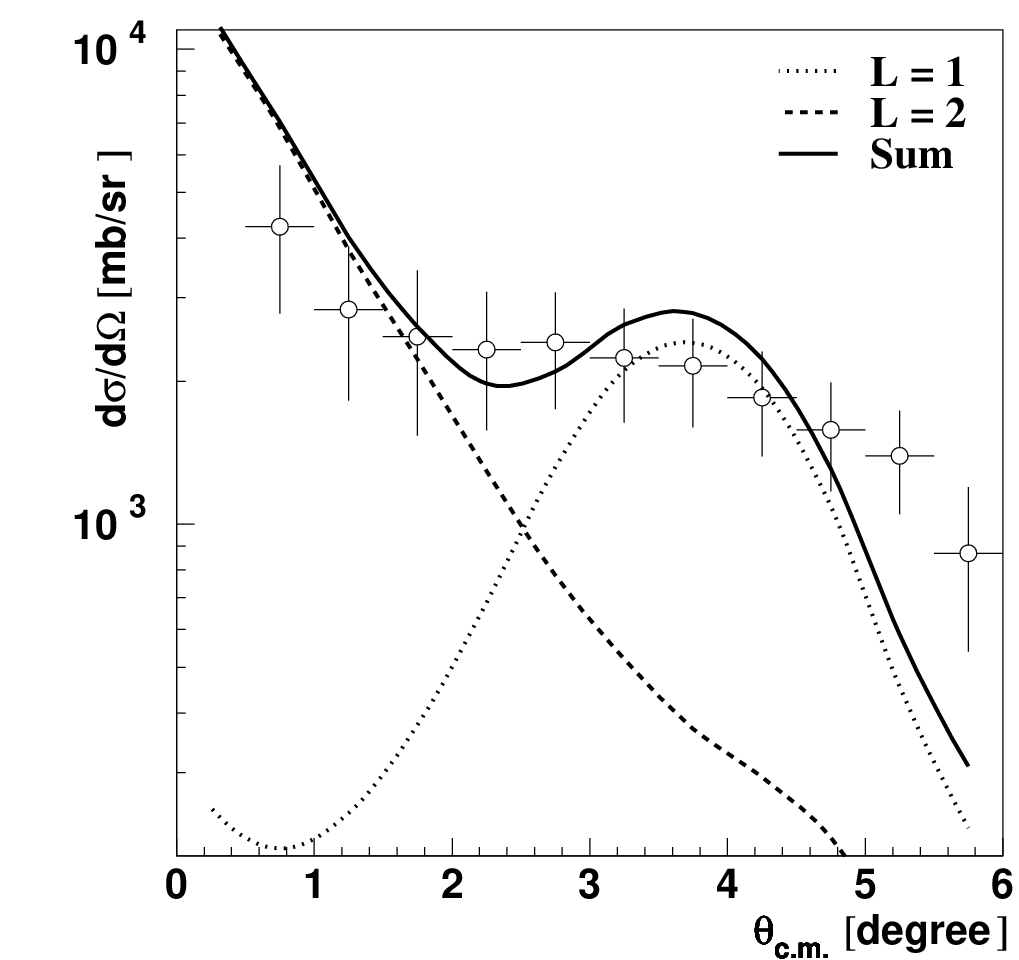}
\caption{Result (solid line) of the multipole decomposition of the
experimental differential cross-section of the structure at
$E^*\sim9$~MeV excited in \Ne{26} using the lead target, and
contributions of $L=1$ (dotted line) and $L=2$ (dashed line) multipoles.
\label{fig:pygmy:angdist:pb:one}}
\end{figure}

The high granularity of the present setup, allows to reconstruct the
scattering angular distribution for \Ne{26} on the \Pb{0} target
(Fig.~\ref{fig:pygmy:angdist:pb:one}) and to extract the E1 excitation
by means of a multipole decomposition analysis. The $L=1$ and $L=2$
angular distributions (dotted and dashed lines) were obtained from
simulations based on {\sc ecis97} angular distribution calculated for
$E^*=9$~MeV. The data were fitted with a linear combination of the two
distributions since $L=2$ and $L>2$ distributions were found to
exhibit similar shapes. The result of the fit gives B(E1) =
0.49$\pm$0.16~\efm{1} which corresponds to 4.9$\pm$1.6\% of the TRK
sum rule around 9~MeV excitation energy. Assuming that the remaining
part of the contribution is due to $L=2$ excitation, we obtain
B(E2$\uparrow$) = 49$\pm$8~\efm{2}.  The two methods to extract the E1
component
thus provide very consistent results. A third method presented in
\cite{gibelin:phd} also leads to the same conclusion. This shows that,
within the error bars, possible contribution of modes such as
isoscalar $L=1$ states to the data taken with the Al target does not
affect the final result on B(E1).


\begingroup
\begin{table*}
\center
\caption[Branching ratios]{Experimental neutron branching ratios for 
  the structure at $E^*\sim9$~MeV in \Ne{26} to the \Ne{25} states, compared 
  to statistical decay calculations  \cite{Puhlhofer77:cascade} for several
  multipolarities. We assumed that the  reaction on \Al{27} induces only 
  $L=2$ transitions, see text.\label{tab:stat:decay}}
\begin{ruledtabular}
\begin{tabular}{cccccccccc}
\multicolumn{3}{c}{Final \Ne{25} state} & 
\multicolumn{4}{c}{Experiment} &
\multicolumn{3}{c}{Statistical decay} \\
\cline{1-2}\cline{4-6}\cline{8-10}
Energy (MeV)& $J^\pi$     & &  Pb            &  Pb $(L=1)$     & Pb $(L=2)=$  Al            & & $L=1$& $L=2$& $L=3$ \\
\hline 		     		           		     		 		    	    		    
0.0         & $1/2^+$     & &$5^{+17}_{-5}\%$ &$5^{+32}_{-5}\%$&   $4^{+5}_{-4}\%$   & & 40\% & 28\% & 22\%\\
1.7 \& 2.0  &$5/2^++3/2^+$& &$66\%\pm15\%$   &$42\%\pm30\%$  &   $95^{+5}_{-15}\%$  & &55\% & 67\% & 75\%\\
3.3         &  $(3/2^-)$  & & $35\%\pm 9\%$  &$60\%\pm17\%$  &     $ 5^{+6}_{-5}\%$   & & 5\%  & 4\% & 3\%\\
\end{tabular} 
\end{ruledtabular}
\end{table*}	
\endgroup

Theoretical calculations have been performed within various
frameworks. Using the relativistic QRPA (RQRPA) and the response
function formalism, Cao and Ma \cite{ma:prc} predict an E1 pygmy state
centered around 8.4 MeV and exhausting 4.5\% of the TRK sum rule,
close to our experimental values. RQRPA calculations for axially
deformed nuclei also report a pygmy state below 10 MeV excitation
energy \cite{arteaga2007rqr}. No corresponding percentage of TRK sum
rule is reported. Similarly, a redistribution of the strength with a
peak at low energy is also predicted by (non-relativistic) deformed
QRPA calculations using Gogny forces \cite{Peru:2007} and Skyrme
forces \cite{Yoshida:2008:arx}. All these calculations agree on the
presence of a structure at low excitation energy in the E1 response
function. In order to get deeper insight into the microscopic
structure of these states, one can examine the dominant configurations
involved, which can be extracted from the calculations of
refs. \cite{Peru:2007,Yoshida:2008:arx,arteaga2007rqr} performed in
the matrix formalism.  In the first two calculations, the dominant
transitions are found to be 2s$_{1/2}\!\to$2p$_{3/2}$ and/or
2s$_{1/2}\!\to$2p$_{1/2}$, corresponding to the promotion of neutrons
essentially from the last occupied orbit of \Ne{26} to fp shells.

It is well-known that the decay pattern of continuum states can give
access to the components of the wave-function of these states.  The
excitation energy reconstruction method used in the present experiment
\cite{gibelin:phd} allows us to extract for the first time data on the
decay of pygmy resonances of neutron-rich nuclei. The experimental
branching ratios to bound states of \Ne{25} are presented in
Table~\ref{tab:stat:decay}. For both Pb and Al targets, the branching
ratio for the decay to the ground-state (g.s.) of \Ne{25} is
compatible with zero. By using the simulation code, it was checked
that any decay pattern including a sizable branch to the g.s. is
incompatible with the experimental results. The large difference
between branching ratios obtained with the two targets proves that
states of different nature have been excited. We then deduced the
$L=1$ and $L=2$ components on the Pb target by assuming that the Al
target induces only a $L \ge 2$ excitation, as described above.  For
comparison, we performed a statistical decay calculation assuming
$L=1,2,3$ emitting states using the {\sc cascade} code
\cite{Puhlhofer77:cascade}, spins and parities of populated states in
\Ne{25} being those listed in Table~\ref{tab:stat:decay}.  The decay
is not statistical, as observed in light nuclei \cite{vdw:book}. Since
the g.s. of \Ne{25} has $J^\pi=1/2^+$, the decay to this state becomes
weaker with increasing spin of the emitting state due to penetrability
effects.

A striking feature of the observed decay pattern is the absence of
decay to the \Ne{25} g.s., which is in contradiction with the
predicted structure of the pygmy states.  Indeed, it is established
that the \Ne{25} g.s. configuration mainly corresponds to a neutron in
the 2s$_{1/2}$ orbit, the experimental spectroscopic factor obtained
from the \Ne{24}(d,p) reaction being 0.8 \cite{Fernandez2007}. If the
main configuration of the pygmy state were actually
$\nu$(2s$_{1/2}^{-1}$2p$_{3/2}$) or $\nu$(2s$_{1/2}^{-1}$2p$_{1/2}$) a
strong decay to the \Ne{25} g.s. should occur, even favoured by
penetrabilities.  This discrepancy indicates that the populated pygmy
states are more mixed and/or involve different
transitions. Interestingly, calculations reported in
\cite{arteaga2007rqr} predict a dominant contribution of the
$K^\pi=1^-$ state, with nearly equal weights of
$\nu$(2s$_{1/2}^{-1}$2p$_{1/2}$) and $\nu$(1d$_{5/2}^{-1}$1f$_{7/2}$)
transitions which is in better qualitative agreement with our data.

Theoretical branching ratios, presently not available, are highly
desirable for a more precise comparison. We note that they could also
be obtained from shell-model calculations by combining the
single-particle spectroscopic factors and penetrability coefficients.


In summary the present study of the neutron rich nucleus \Ne{26} using
intermediate energy inelastic scattering has shown the presence of
pygmy states located around 9 MeV excitation energy. The contribution
of E1 states corresponds to nearly 5\% of the TRK sum rule. These
global features are in agreement with self consistent mean-field
calculations performed in various frameworks.  The decay pattern of
the observed pygmy states has been measured for the first time,
providing a stringent test of the microscopic models describing the
wave function of these states. The measured decay pattern is not
consistent with models predicting a structure corresponding to
excitations of neutrons from the Fermi surface.  Making use of the new
facilities RIBF and Big~RIPS, future studies will investigate nuclei
located even further from stability.


\end{document}